\documentclass[aps,prd,twocolumn,groupedaddress,showpacs]{revtex4}
\newcommand*{\ket}[1]{\ensuremath{|#1\rangle}}
\begin{document}

\title{Effective gauge group of pure loop quantum gravity is $SO(3)$: New estimate of the Immirzi parameter}

\author{Chung-Hsien Chou$^{1}$, Yi Ling$^{2}$, Chopin
Soo$^{3}$ and Hoi-Lai Yu$^{1}$} \affiliation{$^{1}$ Institute of
Physics,
Academia Sinica, Taipei 115, Taiwan\\
$^{2}$ Center for Gravity and Relativity, Department of Physics,
  NanChang University, NanChang 330047, China\\
$^{3}$ Department of Physics, National Cheng Kung University,
Tainan, Taiwan }

%\author{Chung-Hsien Chou}\email{chouch@phys.sinica.edu.tw}
%\affiliation{Institute of Physics, Academia Sinica, Taipei 115,
%Taiwan}
%\author{Yi Ling }\email{yling@ncu.edu.cn}
%  \affiliation{Center for Gravity and Relativity, Department of Physics,
%  NanChang University, NanChang 330047, China}
%\author{Chopin Soo}\email{cpsoo@mail.ncku.edu.tw}
%  \affiliation{Department of Physics, National Cheng Kung University, Tainan, Taiwan}
%\author{Hoi-Lai Yu}\email{hlyu@phys.sinica.edu.tw}
% \affiliation{Institute of Physics, Academia Sinica, Taipei, Taiwan}
%\date{Draft \today}
\begin{abstract}

We argue that the effective gauge group for {\it pure}
four-dimensional loop quantum gravity(LQG) is $SO(3)$ (or
$SO(3,C)$) instead of $SU(2)$ (or $SL(2,C)$). As a result, links
with half-integer spins in spin network states are not realized
for {\it pure} LQG, implying a modification of the spectra of area
and volume operators. Our observations imply a new value of
$\gamma \approx 0.170$ for the Immirzi parameter which is obtained
from matching the Bekenstein-Hawking entropy to the number of
states from LQG calculations. Moreover, even if the dominant
contribution to the entropy is not assumed to come from
configurations with the minimum spins, the results of both pure
LQG and the supersymmetric extension of LQG can be made compatible
when only integer spins are realized for the former, while the
latter also contains half-integer spins, together with an Immirzi
parameter for the supersymmetric case which is twice the value of
the $SO(3)$ theory. We also verify that the $-\frac{1}{2}$
coefficient of logarithmic correction to the Bekenstein-Hawking
entropy formula is robust, independent of whether only integer, or
also half-integer spins, are realized.
\end{abstract}
\pacs{04.60.Ds}
%\keywords{}
\maketitle
\section{Introduction}
The simplification of the constraints and introduction of gauge
variables\cite{A}, and subsequently loop variables, and spin
network states, have brought excitement to the non-perturbative
and background-independent program of canonical quantization of
four-dimensional gravity\cite{RT}. This program is often referred
to as ``loop quantum gravity"(LQG). The construction of loop and
spin network states have moreover proved fruitful, and have
yielded discrete spectra for the well-defined area and volume
operators. In the literature, it is common to encounter the
alternative use of spinorial variables, and to identify the gauge
group of the theory as $SU(2)$ (or its complexification
$SL(2,C)$), although $SO(3)$ (respectively $SO(3,C)$) have the
same $su(2)$ Lie algebra. This is also manifest in the
construction of spin networks and the spectra of area and volume
operators. Explicitly, links in spin networks are usually labeled
with representations of $SU(2)$ with integer and half-integer
spins; consequently the area operator has eigenvalues
$A(j)=8\pi\gamma l_p^2\sqrt{j(j+1)}$, where $l_p$ is the Planck
length and $\gamma$ is the Immirzi parameter\cite{Im} which
reflects a freedom of choice in the theory. However, when the
Immirzi parameter is fixed by comparing, as proposed in
Ref.\cite{DR}, the LQG results for the Bekenstein-Hawking entropy
and the area spectrum to the corresponding quasi-normal mode
calculations of a Schwarzschild black hole with the assumption
that states with the minimum spin are dominant, this matching
yields $j_{min} = 1$, instead of the expected $j_{min} = 1/2$
configurations, indirectly hinting that the gauge group should be
$SO(3)$ instead\cite{DR}. In the supersymmetric extension of LQG,
the same procedure gave ${\tilde j}_{min}= 1/2$ instead and the
value of $\gamma$ which is twice its value for pure LQG\cite{Yi}.
More recent analyses with proper counting of states in
Refs.\cite{dom,Mei,GM,TN} have revealed that the dominance of
minimum spin configurations is both questionable and
underestimates the value of the Immirzi parameter. We shall
discuss in Section IV the related computations and modifications
in black hole physics in the light of these recent analyses and
our conclusions, and shall also confirm that the coefficient of
the area-dependent logarithmic correction to the
Bekenstein-Hawking entropy formula is $-\frac{1}{2}$ and is
independent of the gauge group\cite{LA}.

We shall show that the effective gauge group of pure LQG without
fermions is indeed $SO(3)$ (or its complexification), and is
correspondingly lifted to its covering group $SU(2)$ (or its
complexification) when fermions are present. Our arguments are
based upon 1)retracing the steps which lead to the basic
variables, 2)the fact that a gauge connection transforms according
to the adjoint representation, and 3)the criterion that the
effective gauge group of a theory is determined by its full
physical contents. These observations are rather elementary, but
they impact upon basic calculations in LQG and quantum geometry
with spin network states. The Hilbert space of quantum gravity
without fermionic matter should allow spin network states with
links of {\it integer} $j$ representations only, and links with
half-integer representations do not occur in pure LQG. We would
like to stress that our conclusions on the relevant gauge groups
are neither dependent upon, nor confined to, black hole
situations.
\section{Basic conjugate variables}
We first retrace the steps which lead to the basic variables. In
four, and only four dimensions, the Lorentz algebra can be
decomposed into two $su(2)$ (or $so(3)$) Lie algebras, with self-
and anti-self-dual generators. It follows that finite-dimensional
irreducible fields can be labeled by $(j_+, j_-)$ and transform
according to $(2j_+ +1)\times(2j_- +1)$ representation wherein
$j_\pm$ take positive integer or half-integer values. Self- and
antiself-dual two-forms, $\Sigma^\pm_{a}$, constructed from
vierbein one-forms $e_A =e_{A\mu}dx^\mu$ through $\Sigma^\pm_{a} =
\pm i e_{0}\wedge e_a
+\frac{1}{2}\epsilon_{0a}\,\,^{bc}e_{b}\wedge e_c$ and the
curvature, $F^\pm_a = dA^\pm_a + \frac{1}{2}\epsilon_a\,^{bc}
A^\pm_b\wedge A^\pm_c$, of the self- and anti-self-dual spin
connections, $A^\pm_a = \pm i\omega_{0a} +
\frac{1}{2}\epsilon_{0a}\,^{bc}\omega_{bc}$, are
(1,0)(respectively (0,1)) fields\cite{FN1}.

The covariant Samuel-Jacobson-Smolin
action\cite{Samuel-Jacobson-Smolin} (with cosmological constant
$\lambda$) which yields the Ashtekar variables is\cite{FN2}
 $\frac{-ic^3}{8\pi{G}} {\int} [F^-_{a}\wedge
\Sigma^{-a} +\frac{\lambda}{6}\Sigma^-_a \wedge \Sigma^{-{a}}]$.
To arrive at the canonical variables, we may invoke the ``spatial
gauge"\cite{Kodama} wherein the (real-valued) vierbein and its
inverse assume the form
\begin{equation}
e_{A\mu}=\left[\matrix{ N &0 \cr N^jE_{aj}&
E_{ai}}\right]\quad,\quad e^{\mu{A}} =\left[\matrix{N^{-1}& {0}\cr
-{(N^i/N)}&{E^{ia}}} \right].
\end{equation}
If this is done, the conjugate momentum to $A^-_{ia}$ on the
Cauchy surface with constant $x^0$ is the real-valued densitized
triad ${\tilde E}^{ia} = \sqrt{det(q)}E^{ia}$; and the residual
gauge group consists of real $SO(3)$ rotations of $E^{ia}$ which
leaves the spatial metric invariant. $A^-_a$ remains a complex
field, but its real and imaginary parts transform separately under
spatial rotations and do not get mixed. It is however {\it not
necessary} to invoke the spatial gauge. It turns out that on the
constant-$x^0$ Cauchy surface, $\Sigma^{-a} =
{{\rlap{\lower2ex\hbox{$\,\,\tilde{}$}}{\epsilon_{ijk}}}}\tilde{\cal
E}^{ia}dx^j\wedge dx^k$, with $\tilde{\cal E}^{ia}$ related to
$E^{ia}$ by a $SO(3,C)$ Lorentz transformation\cite{LT}.
Furthermore, $\tilde{\cal E}^{ia}$ is also precisely conjugate to
$A^-_{ia}$, with Poisson bracket $ \{{\tilde{\cal E}}^{ia}({\vec
x},t),A^{-}_{jb}({\vec y},t)\}_{P.B.} =
i(\frac{8\pi{G}}{{c^3}})\delta^i_j \delta^a_b \delta^3({\vec x} -
{\vec y})$. In this context, both conjugate variables are complex,
and full $SO(3,C)$ Lorentz invariance is retained\cite{A,Kodama}.
Starting from the Einstein-Hilbert-Palatini action, and after
solving secondary constraints\cite{A,Kodama,Ash-Bal-Jo}, the
conjugate variables in the spatial gauge is the real pair
$(\tilde{ E}^{ia}, k_{ia})$, with $k_{ia}$ related to the
extrinsic curvature $K_{ij}$ by $k_{ia} = E^{j}_a K_{ij}$. Within
the canonical formalism, it follows that $A_{ia} \equiv -ik_{ia} +
\Gamma_{ia}$ is conjugate to $\tilde{ E}^{ia}$ as well with
$\Gamma_a$ being the torsionless  spin connection compatible with
the triad. Within this context, the gauge group is made up of
$SO(3)$ spatial rotations. In a later development,
Barbero\cite{Ba} suggested that the real connection $A_{ia} \equiv
\gamma k_{ia} + \Gamma_{ia}$ can be used instead to arrive at a
real phase space formulation of General Relativity if $\gamma$,
the Immirzi parameter, is real.
\section{Effective gauge group of LQG}
The above discussions highlight that regardless of which one of
the previous canonical formulations is employed, the gauge group
is $SO(3)$, either the real group of spatial rotations, or its
complexification, the full Lorentz group $SO(3,C)$. It is not
necessary to invoke spinorial variables to arrive at the
fundamental variables, although one can easily convert
anti-self-dual $SO(3)$ indices to $SU(2)$ primed spinor indices by
contracting with Pauli matrices $[\tau^a]_{{\cal A}'}\,^{{\cal
B}'}$. The dimension of the representation spaces is dependent
upon the global structure of the gauge group; and {\it one needs
to examine the full physical contents of the theory to determine
the actual gauge group}. This latter point of view has been
eloquently advocated, and explicitly illustrated, in Ref.\cite{CM}
(for Yang-Mills gauge fields, see, in particular Section 1.4 of
\cite{CM}). To wit, we should be reminded that a Lie algebra
valued connection 1-form always transforms according to the {\it
adjoint representation} of the group as $A' = g A g^{-1} +
igdg^{-1}; \qquad g\in G$.
 The question is whether all $g\in G$ can be regarded as distinct
elements of the group of transformations. Since the center of the
group, C, commutes with all elements of the Lie algebra, it
follows that, $g$, and $g$ multiplied by any element of the
center, has the same effect. Thus as far as gauge potentials are
concerned, the effective gauge group is not G but G/C. If the
theory contains only gauge potentials, the gauge group of the
theory is therefore G/C. For instance in $SU(N)$ pure Yang-Mills
theory, the gauge group is {\it not} $SU(N)$ but
$SU(N)/Z_N$\cite{CM}. In the case of $SU(2)$, $SU(2)/Z_2 = SO(3)$.
The full physical field content of {\it pure} Ashtekar gravity
consists - depending on one's preference of the different
formulations - of one of the conjugate pairs discussed previously.
In all cases, the effective gauge group acting on the variables (a
gauge connection transforming according to the adjoint
representation, and the conjugate momenta of which transforms
covariantly, but with the requirement that physical triads must
remain unchanged under $2\pi$ rotations), is not $SU(2)$ (or its
complexification $SL(2,C)$), but is either $SO(3)$ or $SO(3,C)$.
This result holds even when a cosmological constant and inflaton
or quintessence scalar (Lorentz singlet) fields are incorporated
into the theory. Would this conclusion be different for the
quantum theory with loop variables and spin networks? For the
presently known formulations of the quantum theory, the answer is
in the negative, because the non-integrable phase factor, ${\cal
P}\exp(i{\int^{\vec{y}}_{{\vec{x}}}}_p \,A \,)$, along any path
$p$ which connects two endpoints is invariant under an element of
the center of the group. Explicitly, under a gauge transformation,
in the transformed holonomy element, $g(\vec{y}){\cal
P}\exp(i{\int^{\vec{y}}_{{\vec{x}}}}_p \,A \,)g^{-1}(\vec{x})$,
both $g(\vec{y})$ and $g^{-1}(\vec{x})$ are {\it effectively}
elements of G/C i.e. $SO(3)$ (or $SO(3,C)$). Turning our attention
to spin networks states, the wave function in the connection
representation, $\langle A\ket{\Gamma, \{v\}, \{j\}}$, is such
that for the spin network, $\Gamma$, a non-integrable phase factor
with $A$ in the spin-$j$ representation is associated to each link
which connects two vertices of $\{v\}$. The important point is
that the dynamical degrees of freedom of the theory lies in $A$
(or the holonomy elements). Consequently, unlike $A$ and the
holonomy elements, the vertices do not transform under gauge
transformations. They are instead chosen to carry the right
combination of Wigner symbols to ensure gauge invariance under
transformations of the holonomy link elements. Bearing in mind the
transformation properties of the non-integrable phase factor, it
follows that the gauge group of the quantum theory is still
effectively $SO(3)$ or $SO(3,C)$ in the absence of fermions.

The configuration space is the space of $SO(3)$ gauge connections
modulo the action of $SO(3)$ gauge group, and it is faithfully
parametrized by holonomy elements of $SO(3)$ connections with
integer spin representations, rather than $SU(2)$ holonomies which
include half-integer spin representations. If half-integer
representations are also allowed, the subsequent ``doubling" of
the spectrum of the composite area operator may be an artifact of
the quantization procedure. There can furthermore be obstructions
to the lifting of the $SO(3)$ gauge group to its covering space.
This is perhaps not too serious from the 3-dimensional perspective
since every orientable 3-dimensional manifold is a spin manifold.
Unlike the $SO(3,1)$ Ashtekar-Sen connection, the Barbero-Immirzi
connection cannot be regarded as the pullback of a connection in
four dimensions\cite{SAM}. Thus when restricted to {\it
orientable} 3-manifolds, $SU(2)$ Barbero-Immirzi connections with
non-(anti)self-dual value of $\gamma \neq \pm i$ may appear to be
consistent, but it is not entirely clear if other contradictions
will arise from 4-dimensional considerations. In 4-dimensions
there exists numerous manifolds which are not spin manifolds
 and explicit Einstein manifolds which do not permit spin structures
are known. On the other hand, it can perhaps also be argued that
usual concepts regarding manifolds need not apply to a theory of
quantum geometry. In the absence of further consistency checks and
empirical evidence, the possibility that half-integer spins are
not realized in pure LQG should not be dismissed; and it is also
more straightforward to adhere to the original configuration space
and consider only $SO(3)$ connections and holonomies.

\section{New estimate of the Immirzi parameter}
Assuming the statistical dominance of configurations with the
minimum spin, Dreyer argued that quasi-normal excitations should
be related to the appearance of punctures with spin $j_{min}$. In
order for the Bekenstein-Hawking entropy formula and the
quasi-normal mode result to agree on the same answer for the
Immirzi parameter, $\gamma = \ln 3 /(2\pi\sqrt 2)$, he inferred
that $j_{min}=1$, thereby supporting the case for
$SO(3)$\cite{DR}. However, Dreyer's arguments in favor of $SO(3)$
has lost their cogency in view of the more careful counting of
states carried out in Refs.\cite{dom,Mei,GM,TN}. We shall also
demonstrate it is possible to reconcile $SO(3)$ with the
Bekenstein-Hawking entropy matching, even if the conjecture of
dominance of minimum spin configurations and its association with
quais-normal mode excitations are in error.

Domagala and Lewandowski\cite{dom} showed that the configurations
should be governed by sequences labelled by
\begin{equation} \sum_i \sqrt{|m_i|(|m_i| +1)} \leq a\equiv
\frac{A}{8\pi\gamma l^2_p}, \qquad \sum_i m_i = 0;
\end{equation}
with $m_i \in {-j_i,-j_i +1, ...,j_i}$ and $j_i \in \mathbf{N}/2.$
The correspondence to quasi-normal modes is thus not
straightforward. Meissner demonstrated that the number of states,
$N(a)$, for a given area is therefore given by $N(a) =
\frac{C_M}{\sqrt{4\pi\beta_M a}}e^{2\pi\gamma_M a}$; and the black
hole entropy is consequently\cite{Mei}
\begin{equation} S =\ln N(a) = (\frac{\gamma_M}{\gamma})\frac{A}{
4l^2_p}-\frac{1}{2}\ln(A/l^2_p) +
\ln\frac{C_M}{\sqrt{2\beta_M\gamma}}.\end{equation} By matching
this result to the Bekenstein-Hawking entropy formula for large
black holes, $\gamma_M =\gamma$ is obtained. Moreover, the
numerical value of $\gamma_M$ can be retrieved from the recursion
formula for $N(a)$ which in the large black hole limit is $N(a)=
\sum^\infty_{k=1} 2N(a-\sqrt{k(k+2)/4})$. For large black holes,
the ansatz $N(a)\propto e^{2\pi\gamma_M a}$ therefore yields
$1=\sum^\infty_{k=1}2e^{-2\pi\gamma_M\sqrt{k(k+1)/4}}$, from which
the numerical result $\gamma =\gamma_M \approx 0.238$ is obtained
for the $SU(2)$ theory\cite{Mei}.

Ghosh and Mitra\cite{GM} treat the punctures as distinct and count
both $j$ and $m$ instead of only surface degrees of freedom. This
yields the formula $\sum_{j\in
\mathbf{N}/2}2[(2j+1)/2]e^{-2\pi\gamma\sqrt{j(j+1)}} =1$, together
with an estimate for the number of states as $\ln N(A)\approx
\ln\{(e^{\frac{A}{4l^2_p}})(A/l^2_p)^{-\frac{1}{2}}\}$. The symbol
${\bf [}...{\bf ]}$ in the sum denotes the integer part which
arises from enforcing $m_i \neq 0$ in the degeneracy
factor\cite{TN}. Thus the Immirzi parameter is estimated from the
preceding series as $\gamma \approx 0.262$ for the $SU(2)$ gauge
group\cite{TN}.

Our conclusions in the previous section is that the gauge group of
{\it pure} LQG is $SO(3)$ instead of $SU(2)$. The analyses of
Refs.\cite{dom,Mei,GM,TN} remain essentially valid, but the area
spectrum should be confined to integer $j$ values, and $m_i$
should accordingly be restricted to $m_i \in \mathbf{N}$ instead.
These adjustments imply that in the recursion relation for
$\gamma_M$ and elsewhere in Ref.\cite{Mei}, the values of k should
be restricted to even integers. The expression for the number of
states $N(a) = \frac{C_M}{\sqrt{4\pi\beta_M a}}e^{2\pi\gamma_M a}$
remains the same, but the numerical value of $\gamma_M$ is now
given by $1=  \sum_{k'\in
\mathbf{N}}2e^{-2\pi\gamma_M\sqrt{k'(k'+1)}}$, giving
$\gamma_{SO(3)} =\gamma_M \approx 0.138$ for the $SO(3)$ theory.
Similarly, the improved value of $\gamma$ following
Refs.\cite{GM,TN} should now be obtained from $\sum_{j\in
\mathbf{N}}2[(2j+1)/2]e^{-2\pi\gamma\sqrt{j(j+1)}} = 1$. These
considerations yield the new value of $\gamma_{SO(3)} \approx
0.170$. The form of the expression for $N(a)$ and the dependence
on the $\ln A$ term in $\ln N(a)$ are unaffected by whether only
integer, or half-integer spins are also allowed, so the
coefficient of logarithmic correction to the Bekenstein-Hawking
formula is thus robust and remains $-1/2$.

What happens when fermions are incorporated into the theory? From
the point of view of the classical fundamental field variables,
although the effective group of the gauge connection remains
$SO(3)$, fermions are not invariant under the non-trivial element
of the center (which corresponds to $2\pi$ rotations) of $SU(2)$,
so now $SU(2)$ (or its complexification $SL(2,C)$) becomes the
effective gauge group of the full physical contents of the theory.
Analogously, the presence of quarks in Quantum Chromodynamics
lifts the physical gauge group from $SU(3)/Z_3$ for the theory of
pure gluons to the full $SU(3)$\cite{CM}. With fermions,
half-integer-spin representations can be effectively realized in
spin network states, as the wave functions are now $\langle A,
\psi\ket{\Psi}$. An example of a spin network with fermionic
degrees of freedom and $SU(2)$ effective gauge group is one with
vertices such that every link with $j=1/2$ has a spin 1/2 fermion
at one of its ends and a vertex(now with $j=1/2$ allowed) with the
correct Wigner symbols at the other; and links with $j\in
\mathbf{N}$ are joint only to other vertices at both ends.

The relationship between black hole entropy based upon loop
quantization of $N=1$ supergravity and quasi-normal mode
excitations based upon the minimum spin contribution was studied
in Ref.\cite{Yi}. In the presence of supergauge fields with
bosonic and fermionic degrees of freedom, the effective gauge
group is lifted to the covering group. The area spectrum for a
link of spin $\tilde j$ has been calculated to be
$A_{SUGRA}(\tilde{j}) = 8\pi{\tilde\gamma}l_p^2\sqrt{\tilde
{j}(\tilde{j} + \frac{1}{2})} =
8\pi(\frac{\tilde\gamma}{2})l^2_p\sqrt{j(j + 1)}$, wherein the
non-trivial values are for $\tilde{j} \in \mathbf{N}/2$, or $j\in
\mathbf{N}$;
%A_{SUGRA}(\tilde{j}) &=& 8\pi{\tilde\gamma}l_p^2\sqrt{\tilde
%{j}(\tilde{j} + \frac{1}{2})},\qquad \tilde{j} \in \mathbf{N}/2
%\cr &=& 8\pi(\frac{\tilde\gamma}{2})l^2_p\sqrt{j(j + 1)},\qquad
%j\in \mathbf{N};
%\end{eqnarray}
and the degeneracy of states of a puncture of spin $\tilde j$ is
$(4{\tilde j}+1)$. Thus following the analysis of
Refs.\cite{GM,TN}, the Immirzi parameter now obeys the modified
equation
\begin{eqnarray}1&=&\sum_{{\tilde j}\in
\mathbf{N}/2}2[(4{\tilde
j}+1)/2]e^{-2\pi{\tilde\gamma}\sqrt{{\tilde j}({\tilde
j}+\frac{1}{2})}}\cr &=& \sum_{j \in \mathbf{N}}2[(2j
+1)/2]e^{-2\pi({\tilde\gamma}/2)\sqrt{j(j+1)}}.
\end{eqnarray}
This indicates that the result for the supersymmetric case will be
the same as for the case of $SO(3)$, but with the Immirzi
parameter ${\tilde\gamma}$ for $N=1$ supergravity theory replaced
by $\tilde\gamma = 2\gamma_{SO(3)} \approx 2(0.170)$, despite the
fact that for the supersymmetric case half-integer spins are also
allowed while for pure LQG only integer spins are realized. Note
that this relation between the Immirzi parameters produces exactly
the same area spectrum for {\it both} pure LQG without
supersymmetry and its supersymmetric extension. It also acts as an
intriguing consistency condition, since results for both cases
were calculated and compared for the same black hole masses and
thus same horizon areas of all sizes.
\begin{acknowledgments}
The research for this work has been supported by funds from the
National Science Council of Taiwan under Grant Nos.
NSC93-2112-M-006-011 and NSC94-2112-M-006-006; and the National
Center for Theoretical Science. Y.L. was supported by NSFC
(No.10405027, 10205002) and SRF for ROCS, SEM.
\end{acknowledgments}


\begin{thebibliography}{}
\bibitem{FN1}Lorentz indices are denoted
by Latin letters at the beginning of the alphabet; with uppercase
letters having values from 0 to 3, while lower case letters from a
to h run from 1 to 3. Space-time indices are denoted by Greek
letters, while spatial indices on the Cauchy surface are denoted
by Latin letters from i onwards. Lorentz indices are raised and
lowered by the Minkowski metric which is diag$(-1, 1,1,1)$.
\bibitem{FN2}
Although only one set is required, both $\pm$ sets work equally
well for the purpose of describing pure gravity theory. The choice
of anti-self-dual variables is motivated by the convention that
anti-self-dual spin connections couple to left-handed Weyl
fermions of the Standard Model. For a more complete discussion of
related issues, see, for instance, Ref.\cite{CP}.
\end{thebibliography}
\end{document}